\documentclass[12pt]{iopart}
\usepackage{here}
\usepackage{mathrsfs}
\usepackage{cite}
\usepackage{graphicx}

\newcommand{\fett}[1]{\mbox{\boldmath$#1$}}

\begin{document}

\title[Molecular Complete-Graph Tensor Network]{Complete-Graph 
Tensor Network States: A New Fermionic Wave Function Ansatz for Molecules}

\author{
Konrad H Marti$^1$,
Bela Bauer$^2$,
Markus Reiher$^1$\footnote{Author to whom correspondence should be sent; 
email: markus.reiher@phys.chem.ethz.ch, TEL: ++41-44-63-34308, FAX: ++41-44-63-31594},
Matthias Troyer$^2$ and 
Frank Verstraete$^3$
}

\address{$^1$ Laboratory for Physical Chemistry, ETH Zurich,
Wolfgang-Pauli-Str.\ 10, CH-8093 Zurich}

\address{$^2$ Institute for Theoretical Physics, ETH Zurich,
Wolfgang-Pauli-Str.\ 27, CH-8093 Zurich}

\address{$^3$ Faculty of Physics, University of Vienna,
Boltzmanngasse 5, A-1090 Vienna}

\ead{markus.reiher@phys.chem.ethz.ch}

\begin{abstract}
We present a new class of tensor network states that are
specifically designed to capture the electron correlation of a molecule of arbitrary structure.
In this ansatz, the electronic wave function is represented by a Complete-Graph Tensor 
Network (CGTN) ansatz which implements an efficient reduction of the
number of variational parameters by breaking down the
complexity of the high-dimensional coefficient tensor of a full-configuration-interaction (FCI) wave function. 
We demonstrate that CGTN states approximate ground states of
molecules accurately by comparison of the CGTN and FCI expansion coefficients.
The CGTN parametrization is not biased towards any reference configuration in contrast to many standard quantum chemical methods.
This feature allows one to obtain accurate relative energies between CGTN states which is central to molecular physics and chemistry.
We discuss the implications for quantum chemistry and focus on the spin-state problem.
Our CGTN approach is applied to the energy splitting of states of different spin for methylene and the strongly
correlated ozone molecule at a transition state structure.
The parameters of the tensor network ansatz are variationally optimized by means
of a parallel-tempering Monte Carlo algorithm.
\end{abstract}

{\bf final version: April, 28, 2010}
\maketitle

\section{Introduction}

Over the last two decades, we have witnessed the rise of the Density Matrix
Renormalization Group (DMRG) algorithm \cite{whit92a,whit92b,whit93}, which had a tremendous impact on the
fields of condensed matter physics \cite{scholl05} and quantum chemistry \cite{chan08,mart09}.
In 1995, Rommer and {\"O}stlund showed that the DMRG wave function
\cite{rommerostlund1995,rommerostlund1997} can be
described by a Matrix Product State (MPS) \cite{fannes1992,fannes1992b} which is -- qualitatively speaking --
a one-dimensional chain of rank-3
tensors.
The understanding of the structure of the DMRG wave function
stimulated further developments to efficiently represent ground states
of strongly correlated systems. 

The work of Affleck, Kennedy, Lieb and Tasaki~\cite{aklt} on finitely correlated
states provided the foundations of a new family of states, the tensor network states. 
The basic idea of tensor network states is to approximate ground-state wave
functions of strongly correlated systems by spanning only the relevant part of the Hilbert
space of the system of interest \cite{verstraete2008}.
In the case of a limited amount of entanglement in the system, only a subspace of the full Hilbert space needs to be considered.
This low-entanglement subspace can then be efficiently approximated by tensor network
states tailored to represent the entanglement structure of the system.

In this article, we study a new class of approximations, which we denote Complete-Graph
Tensor Network (CGTN) states, to represent electronic wave functions of
molecular systems described by a complete pair-entanglement network of one-particle states (molecular orbitals). 
A CGTN state provides an efficient and compact description in terms of 
variational parameters because the 2$^k$ expansion parameters for the many-electron states 
that can be constructed from $k$ spin orbitals are approximated by the entries of a matrix symmetric in the
orbital indices, i.e., only $[(k^2-k)/2+k]\times q^2=k(k+1)/2\times q^2$ parameters are needed in our CGTN approach
(where $q$ is the dimension of the one-particle Hilbert space, i.e., $q$=2 for spin orbitals).
Of course, a detailed numerical analysis of the accuracy 
to approximate a total electronic state is required for this reduced parameter
set.
Note that an artificial one-dimensional ordering of the molecular orbitals for the construction of the
total basis states, which can create convergence problems to 
local energy minima as in the quantum chemical DMRG algorithm
\cite{reih05a,reih06}, is explicitly circumvented.

This article is organized as follows:
In Section~\ref{states}, we give a detailed introduction into the application of
tensor network states for molecules. Afterwards, the CGTN ansatz is
described in Section~\ref{cgtnansatz}. In Section~\ref{optimization}, the
procedure for the variational optimization of the CGTN via 
parallel-tempering Monte Carlo is presented. The vertical spin splittings of
methylene and of ozone are reported in Section~\ref{results}.

\section{Novel Representations of Quantum-Many Body States\label{states}}
The electronic Hamiltonian in second quantization reads in Hartree atomic units ('$\hbar=m_e=e=4\pi\epsilon_0=1$')
\begin{equation}
  \label{hamiltonian}
  \hat{H}_{el} = \sum_{i,j \atop \sigma} h_{ij} a^{\dagger}_{i\sigma} a^{}_{j\sigma} +
      \frac{1}{2}\sum_{{i,j,k,l} \atop {\sigma,\sigma'}} V_{ijkl}  a^{\dagger}_{i\sigma} a^{\dagger}_{j\sigma'} a^{}_{k\sigma'} a^{}_{l\sigma} ,
\end{equation}
which contains one-electron integrals $h_{ij}$ over spatial orbitals $\phi_i(\fett{r})$ given in non-relativistic 
theory by \cite{meinbuch}
\begin{equation}
h_{ij} = \int \phi_i^*(\fett{r}) \left( -\frac{1}{2} \nabla^2 - \sum_I
\frac{Z_I}{r_I}\right) \phi_j(\fett{r}) \, d^3r ,
\end{equation}
with nuclear charge number $Z_I$ of atomic nucleus $I$ and electron--nucleus-$I$ distance $r_I=|\fett{r}-\fett{R}_I|$.
The nucleus--nucleus repulsion term is suppressed for the sake of brevity.
The two-electron integrals $V_{ijkl}$ are defined as 
\begin{equation}
V_{ijkl} = \int \! \int \frac{\phi_i^*({\fett{r}}_1) \phi_j^*({\fett{r}}_2) \phi_k({\fett{r}}_2) \phi_l({\fett{r}}_1)}{r_{12}} \, d^3r_1 \, d^3r_2 .
\end{equation}
The Hamiltonian and its ingredients may also be written in terms of spin orbitals $\phi_i(\fett{x})=\phi_i(\fett{r})\sigma$,
where $\sigma$ is a spin-up or spin-down spin eigenfunction.
Coordinate $\fett{x}$ then denotes both spatial and spin variables.

The eigenstate of the electronic Hamiltonian is the total electronic wave function $\vert{\Psi^{(N)}_{A}}\rangle$ if we restrict ourselves
to $N$ electrons.
In quantum chemistry, the full-configuration-interaction (FCI) expansion of the total electronic wave function
in terms of spin-adapted configuration state functions (SU(2) eigenfunctions) exactly solves the non-relativistic 
time-independent Schr\"odinger equation in a given one-particle basis of orbitals.
The FCI wave function can be understood as a linear combination of occupation number vectors in the
direct-product basis of the one-particle Hilbert spaces. Occupation number vectors are
generated by distributing $N$ electrons among the $k$ orbitals. 
The FCI wave function of total electronic state $A$ then reads
\begin{equation}
\vert{\Psi^{(N)}_{A,{\rm FCI}}}\rangle = \sum_{n_1 n_2 \ldots n_k}^q 
C^{(A)}_{n_1 n_2 \ldots n_k} \quad \vert{n_1 n_2 \ldots n_k}\rangle
\label{fci}
\end{equation}
where $C^{(A)}_{n_1 n_2 \ldots n_k}$ are the (F)CI expansion coefficients of 
state $A$ and $\vert{n_1 n_2 \ldots n_k}\rangle$ denotes an occupation number vector.
Note that we restrict the sum in (\ref{fci}) to those vectors that represent $N$ electrons
and thus span the $N$-particle Hilbert space.
The sums run over the dimension $q$ of the local Hilbert spaces of the set
of orbitals $\{ 1, 2, \ldots,k \}$. 
For spin orbitals holds $q=2$, the occupied and unoccupied one-electron state $\{\vert{1}\rangle, \vert{0}\rangle\}$. 
In the case of spatial orbitals, the basis of the local Hilbert space $\{\vert{n_i}\rangle \}$ consists of four states,
$\{ \vert{}\rangle, \vert{\uparrow}\rangle, \vert{\downarrow}\rangle, \vert{\uparrow \downarrow}\rangle \}$. 
Each occupation number vector $\vert n\rangle\equiv\vert{n_1 n_2 \ldots n_k}\rangle$ is
constructed as a direct product from the states of the local Hilbert spaces 
\begin{equation}
\vert n\rangle\equiv\vert{n_1 n_2 \ldots n_k}\rangle = \vert{n_1}\rangle \otimes \vert{n_2}\rangle
\otimes \ldots \otimes \vert{n_k}\rangle.
\label{directproductstructure}
\end{equation}
The number of variational CI parameters required to describe an electronic state 
(or a quantum state in general) grows exponentially with system size, which 
is a direct consequence of the underlying tensor-product structure of the Hilbert space. 

The exponentially growing number of parameters in the FCI ansatz 
renders it unfeasible for molecules containing more than a few atoms. 
Nevertheless, because of the nature of the interactions we may hope that there exists an efficient parametrization of a class of
variational wave functions such that the low-energy sector of the electronic
Hamiltonian is described with sufficient accuracy
\cite{verstraete2008,verstraete2009,vidal2003a,eisert2008,hastings2006}.
In addition, the huge body of numerical evidence compiled during the past forty to fifty years in
quantum chemistry demonstrates that various truncated configuration-interaction expansions are efficient
and reliable to approximate an electronic state \cite{helgaker_book}.
This latter observation indicates that provided we find an efficient
parametrization of all CI coefficients in the FCI expansion, we do not need to
sample all occupation number vectors but only the most important ones for the
total electronic energy. Otherwise, the procedure
would be as unfeasible for large molecules as the FCI approach itself is.

A way of finding an efficient parametrization of states is to approximate the
high-dimensional coefficient tensor $C^{(A)}_{n_1 n_2 \ldots n_k}$ by a
\emph{tensor network}.
Tensor network states build a new class of variational wave functions. The high-dimensional coefficient tensor is broken
down into low-rank tensors which are arranged on an arbitrary
network \cite{sierra1998,nishino2001,nishio2004,verstraete2004,martindelgado2002,verstraete2006,zhou2008,verstraete2008}. 
The primary advantage of tensor network states compared to the standard FCI expansion is the
reduced number of variational parameters which approximately scales as $O(k
\chi^p)$ where $k$ is the number of orbitals, $\chi$ the bond dimension and $p$
is the rank of the tensor. Tensor network states can be designed in
a way to directly map the entanglement of the underlying system~\cite{shi2006,vidal2007}.

The MPS constructed by the DMRG algorithm are the simplest example of tensor network states for one-dimensional systems
\cite{verstraete2004,verstraete2006,perez2007}. 
An MPS with open-boundary conditions is defined as
\begin{equation}
\vert{\Psi^{(N)}_{\rm MPS}}\rangle = \sum_{n_1 n_2 \ldots n_k} A_{1}[n_1] A_{2}[n_2]
\cdots A_{k}[n_k] \quad \vert{n_1 n_2 \ldots n_k}\rangle
\label{mpsstate}
\end{equation}
where the rank-3 tensors $A_i$ are written as $m \times m$ matrices $A_i[n_i]$ for a specific
local state $n_i$~\cite{klumper1991,klumper1992,fannes1992,derrida1993}. 
Note that $A_1[n_1]$ and $A_{k}[n_k]$ are vectors because open-boundary conditions are
applied and that we have dropped here and in the following the state index $A$ for the sake of brevity. 
The DMRG algorithm optimizes the tensors by keeping the eigenvectors of the
reduced density matrix corresponding to the dominant eigenvalues
(see also McCulloch \cite{mccul2007} and Verstraete~\emph{et al.}
 \cite{verstraete2008} who discussed the additional flexibility
when using both wave function and Hamiltonian in MPS representation).
Chan~\emph{et al.} and Hachmann~\emph{et al.} rephrased the quantum chemical
DMRG algorithm consistently in terms of matrix product
states~\cite{chan08,chan06}. 

Other variational families of states have been proposed for strongly correlated systems which can be
efficiently contracted for Variational Monte Carlo calculations. These include
string-bond states (SBS) \cite{schuch2008} and subsequently
Entangled-Plaquette States (EPS)~\cite{mezzacapo2009} and Correlator Product States (CPS)~\cite{chan09}. 
In this work, we will build upon an extension of these families to treat the full
electronic Hamiltonian for molecular systems. Our extension is twofold: (i) SBS, EPS and CPS have only
been applied to local spin Hamiltonians so far, while we aim at the full electronic Hamiltonian
as given in (\ref{hamiltonian}) and (ii) we include all pair correlations of the one-electron basis states
and do not impose any restriction on these pairs.

\section{Complete-Graph Tensor Network Ansatz\label{cgtnansatz}}

In the case of an MPS parametrization of a wave function, sites --- or orbitals in a quantum chemical
context --- have to be mapped on a suitably chosen lattice.
Then, correlations are transmitted over the one-dimensional lattice by the size of the matrices occurring in the matrix product state. 
Naturally, this ansatz is more suitable for molecular systems with an inherent linear structure rather than for 
those with long-range correlations.
Orbital ordering on this lattice is then crucial for the convergence of the
variational optimization technique employed, e.g., for the DMRG algorithm~\cite{reih05a,reih06}. 
Hence, an MPS state might be difficult to optimize for a general molecule of arbitrary structure.
By contrast, in the CGTN approach to be introduced now non-local correlations are directly embedded into the
non-linear tensor network ansatz. The Complete-Graph Tensor Network replaces the high-dimensional coefficient
tensor in the FCI ansatz of (\ref{fci}) by a network of tensors that connects all orbitals with each other,
\begin{equation}
\vert{ \Psi^{(N)}_{\rm CGTN} }\rangle = \sum_{n_1 n_2 \ldots n_k}^q \prod_\alpha^k \prod_{\beta
\leq \alpha}
f_{\alpha \beta}^{n_\alpha n_\beta} 
\quad 
\vert{n_1 n_2 \ldots n_k}\rangle
\label{ansatz}
\end{equation}
where $\fett{f}\equiv\{f^{n_\alpha n_\beta}_{\alpha \beta}\}$ represents a rank-[$k(k+1)/2$] tensor which depends on the orbitals $\alpha, \beta \in
\{1,2,..,k\}$. 
The local states of the spin orbitals $n_\alpha, n_\beta$ can either be
occupied or unoccupied $\{ \vert{1}\rangle, \vert{0}\rangle \}$ (i.e., for spin orbitals $q$=2).
The sum runs over all possible occupation number vectors $\vert{n}\rangle$ in the $N$-electron Hilbert space
(in principle, in the full Fock space)
with the correct number of electrons, projected spin, and point-group symmetry.

The above ansatz is built on the key idea that every orbital is ``connected'' with every other orbital.  
Hence, all CI coefficients are constructed from such pair correlations optimized for {\it all} orbitals.
The number of variational parameters in our ansatz depends on the number of
spin orbitals $k$ and on the bond dimension $d$ and is given as $\frac{1}{2} k(k+1) q^2 $
where $d=2$. Comparing this to the number of parameters in the FCI ansatz which
scales as $O(2^k)$ for spin orbitals, it is clear that CGTN states are much
more efficient in terms of the number of variational parameters. 
It is important to emphasize that we do not need to specify any reference 
configuration like in most post-Hartree--Fock methods. Our ansatz comprises naturally all
basis states which can be generated in the Hilbert space of interest.
Thus, although the CI coefficients are approximated by the reduced set of CGTN
parameters, we can expect that CGTN calculations are (at least approximately) size-consistent.

Compared to the tensor networks suggested so far for local (spin) Hamiltonians (see last section), 
CGTN states form a subspace of the very general CPS parametrization, which is so general that it basically covers any 
non-hierarchical tensor network approximation of a wave function. 
In particular, they correspond to two-site CPS including all long-range correlation effects. A similar parametrization
of a simple variational wave function was also chosen by Huse and Else to describe ground states
of frustrated quantum spin systems~\cite{huse1988}. However, the question arises how accurate this parametrization
is for the non-local electronic Hamiltonian of (\ref{hamiltonian}), which shall be investigated in this work.
Although undesirable from the point of view of feasibility, inaccuracies may be cured by also including three-orbital,
four-orbital, ... correlations as is possible with the 
general CPS ansatz. Thus, we may easily increase the flexibility of CGTN states by
including higher-order correlators (summing over three or more indices instead
of two) or by increasing the bond dimension of the tensors from scalar values to matrices.
In this work, the number of pair-correlation parameters is determined by the definition of an active orbital space,
which is a standard procedure in quantum chemistry \cite{helgaker_book}.
The next step is to variationally optimize the non-linear tensor network ansatz.

\section{Variational Optimization\label{optimization}}

We apply a variational Monte Carlo scheme to optimize the CGTN state. 
In the context of tensor-network states, this was demonstrated by
Schuch~\emph{et al.}~\cite{schuch2008} and by Sandvik and Vidal~\cite{sandvik2007} for local Hamiltonians. 
We augment the optimization with a parallel tempering scheme.
The energy of the system is written as an expectation value of the Hamiltonian
operator over an $N$-electron wave function $\vert\Psi^{(N)}_{\rm CGTN}\rangle$,
\begin{equation}
E_{\rm{FCI}} = 
\frac{\langle{\Psi^{(N)}_{\rm{FCI}}}\vert{\hat
H_{el}}\vert{\Psi^{(N)}_{\rm{FCI}}}\rangle}{\langle{\Psi}^{(N)}_{\rm{FCI}}\vert{\Psi^{(N)}_{\rm{FCI}}}\rangle} \leq
\frac{\langle{\Psi^{(N)}_{\rm CGTN}}\vert{\hat
H_{el}}\vert{\Psi^{(N)}_{\rm CGTN}}\rangle}{\langle{\Psi^{(N)}_{\rm CGTN}}\vert{\Psi^{(N)}_{\rm CGTN}}\rangle}
\label{variationalenergy}
\end{equation}
which delivers an upper bound to the exact FCI energy in a given one-particle basis.

For the sake of brevity, the tensor product in front of the occupation number vector in our CGTN
ansatz is abbreviated by a scalar function $C_I$,
\begin{equation}
C_I = \langle{I}\vert{\Psi^{(N)}_{\rm CGTN}}\rangle =
\prod_\alpha^k \prod_{\beta \leq \alpha}
f_{\alpha \beta}^{I_\alpha I_\beta} 
\end{equation}
which corresponds to a (unnormalized) CI-like coefficient
for a given occupation number vector $\vert{I}\rangle$ in configuration-interaction theory,
\begin{equation}
\vert{\Psi^{(N)}_{\rm CGTN}}\rangle = \sum_I C_I \vert{I}\rangle 
\label{weightansatz}
\end{equation}
Inserting (\ref{weightansatz}) into the normalization integral in the denominator of (\ref{variationalenergy}) 
yields the well-known CI-like normalization condition
\begin{equation}
\langle{\Psi^{(N)}_{\rm CGTN}}\vert{\Psi^{(N)}_{\rm CGTN}}\rangle =
\sum_{I K} C^\star_{I} C_{K} \langle{I}\vert{K}\rangle =
\sum_{I K} C^\star_{I} C_{K} \delta_{I K} =
\sum_{I} C^2_{I} 
\label{normalization}
\end{equation}
where the sum takes the square of the weights over all possible basis states in the Hilbert space with correct symmetry.

We now insert (\ref{normalization}) into (\ref{variationalenergy})
to get an approximation to the energy expectation value of the electronic Hamiltonian for our wave function
ansatz and then substitute the ket in the denominator by (\ref{weightansatz}),
\begin{equation}
E_{\rm{approx}} = \frac{ \sum_I C_I \langle{\Psi^{(N)}_{\rm CGTN}}\vert{\hat
H_{el}}\vert{ I }\rangle}{ \sum_{I} C^2_{I} } .
\label{mod1}
\end{equation}
After rewriting this sum to become
\begin{equation}
E_{\rm{approx}} = \frac{ \displaystyle \sum_I C^2_I
\displaystyle \frac{\langle{\Psi^{(N)}_{\rm CGTN}}\vert{\hat H}_{el}\vert{ I }\rangle}{C_I } }
{\displaystyle \sum_{I} C^2_{I} }
\label{mod1}
\end{equation}
we can perform Monte Carlo sampling with strictly non-negative probabilities $C_I^2$.
We define an energy estimator $E(I)$ as a function of the
occupation number vector $\vert{I}\rangle$ that reads
\begin{equation}
E(I) \equiv
\frac{\langle{\Psi^{(N)}_{\rm CGTN}}\vert{\hat H}_{el}\vert{ I }\rangle}{C_I} =
\sum_{J} \frac{C_{J}}{C_I} \langle{J}\vert{\hat H}_{el}\vert{ I }\rangle
\label{energyestimator}
\end{equation}

For a given $|I\rangle$, the number of basis states $\langle J|$ contributing to this sum is 
bounded by the number of terms in the Hamiltonian. 
Since the occupation number vector $\vert I\rangle$ is
not an eigenstate of the Hamiltonian, $\hat{H}_{el}\vert{I}\rangle$ produces a
linear combination of occupation number vectors with coefficients
constructed from the one-electron and two-electron integrals occurring in the
Hamiltonian. For a molecule, the sum over $J$ can therefore be performed exactly.

A variational optimization is in general not guaranteed to converge to a global minimum --- instead, it may be trapped in local minima 
and yield incorrect results. In our ansatz, the highly non-linear structure and the complex energy landscape of molecules make such 
trapping quite likely. In particular, the approach of Sandvik and Vidal~\cite{sandvik2007} which applies gradient information to 
propose a new set of variational parameters turns out to be unreliable in our case.
We therefore use a stochastic optimization procedure that works entirely without gradient information. 
To each choice of variational parameters $\fett{f}=\{f^{n_\alpha n_\beta}_{\alpha \beta}\}$, we can assign an electronic energy $E(\fett{f})$. 
After introducing an artificial temperature $T$ (actually, a parameter with the dimension of an
energy; here measured in Hartree units, $E_h$), we can sample the continuous variables $\fett{f}$ following a canonical ensemble 
with the weight of a configuration given by $\exp( - E(\fett{f})/T)$. The limit
$T$$\rightarrow$0 $E_h$ will yield the desired ground 
state of the molecule. It should be emphasized that this ensemble does not correspond to a physical ensemble at any finite temperature.

The advantage of this approach is that we can easily control the optimization
procedure by tuning the parameter $T$. 
While an accurate simulation of the ground state is only possible for $T$$\rightarrow$0 $E_h$ that may get stuck in local minima, a simulation 
at larger $T$ may easily surmount high energy barriers between local minima. We therefore use a parallel tempering/replica exchange scheme \cite{fren02}
where simulations are run at several temperatures simultaneously. After a certain number of
updates, replica $i$ and replica $i+1$ at neighbouring temperatures are exchanged with a probability
\begin{equation}
p\left( (T_i, E_i) \leftrightarrow (T_{i+1}, E_{i+1}) \right) = \min\{1,\exp\left( -\Delta E / \Delta T \right)\}
\end{equation}
with $\Delta E=E_{i+1}-E_{i}$ and $\Delta T=T_{i+1}T_{i}/(T_{i}-T_{i+1})$.
The set of temperatures has to be chosen in such a way that the lowest
temperatures are close to $T$=0 $E_h$ to yield information about the 
ground state and the highest temperatures are sufficient to overcome energy barriers. Hence, the choice depends very much on the specific 
problem at hand. 
It might be desirable to dynamically optimize the temperatures for some specific
applications~\cite{katzgraber2006}, but for the purpose of this work, we restrict ourselves to a
static choice of the temperatures.
For a temperature set of $M$ temperatures in the range
$[T_{1}, T_{M}]$, we choose for the remaining $M-2$ temperatures $T_l$ with $T_1<T_l<T_M$:
\begin{equation}
T_l = T_{1} \left( \exp {\frac{\ln T_{M} - \ln T_{1}}{M-1}} \right)^{l-1},\ l = 1 \ldots M.
\end{equation}
It is, of course, possible to finally take the state obtained from the above procedure as an input state for a direct optimization using gradient 
information, which may yield better accuracy close to the minimum.

\section{Results\label{results}}
Our primary goals in this work are: (i) to analyze the CGTN ansatz for the description of
electronic energies and CI coefficients
of molecules and (ii) to show that we can optimize the CGTN ansatz
by means of a variational parallel-tempering Monte Carlo algorithm. 
Our test molecules are methylene and ozone. For these small molecules
we do not need to apply sampling of the occupation number vectors since 
the sum in (\ref{mod1}) can be carried out explicitly.
Hence, we use the above-described sampling scheme to sample the $\fett{f}$ 
coefficients only. As a consequence, we avoid the sampling error of the occupation number vectors
and thus obtain a reliable picture of the quality of our CGTN ansatz.

\subsection{Singlet and Triplet Polyatomic Radicals: The Methylene Example\label{ch2}}
The accurate calculation of different spin states is of great importance to
chemistry; in particular, for chemical reactions in which a spin-crossing event
occurs~\cite{paul01,reih01,reih02h,salo02,poli03,harv04,reih06_fd135,schr00,shaik1999}.
There is, however, no way to tell our optimization algorithm how to converge
directly to the desired spin state. The optimization algorithm might get easily trapped in local
minima corresponding to a spin-contaminated total state.
One possible solution would be to sample over the basis of spin-adapted configuration state
functions (CSF) which can easily be constructed as linear combinations of
occupation number representations using Clebsch--Gordon coefficients
producing a SU(2) eigenstate of the Hamiltonian with a well-defined total spin. 
In that case, however, 
the complete occupation number vector basis must be constructed.
Another solution, which we need to employ in our second example below, is the application of a
level-shift operator as used, for instance, for the DMRG algorithm in \cite{marti2008}.
The concept of the level-shift operator can be easily implemented in the current
optimization scheme. The idea is to substitute the original Hamiltonian by an
effective shifted Hamiltonian where the unwanted states with a higher
multiplicity are shifted up in energy. The lowest energy state of the total system is then
a spin-pure state with the correct total spin.
The shifted Hamiltonian is written as
\begin{equation}
\hat{H}_{shifted} = \hat{H} + \epsilon \hat{S}_- \hat{S}_+ = H + \epsilon \left( \hat{S}^2
- \hat{S}^2_z - \hat{S}_z \right)
\end{equation}
where we add the product of the spin ladder operator to the original Hamiltonian
and $\epsilon$ is a positive constant.
This prevents the occurrence of states which possess the same projected spin but
have a different total spin.

We choose methylene as our test molecule, for which we determine the spin splitting of the
singlet and triplet states. Methylene is the smallest polyatomic radical featuring a triplet ground state and several low-lying singlet states where
strong correlations effects are present~\cite{kalemos04}.
We are particularly interested in the energies of the triplet ground state and
the lowest-lying
singlet state with point-group symmetries $B_1$ and $A_1$, respectively.
 
In preparatory calculations, we calculated the one-electron and two-electron integrals
as well as complete-active-space (CAS) self-consistent-field (SCF) reference energies with the {\sc Molpro} program
package~\cite{molpro2002.6}. The orbitals have been expanded in Dunning's cc-pVTZ basis set~\cite{dunn1989,dunn1992}.
The electronic energies of the singlet and triplet state of CH$_2$ were studied at a C--H bond
distance of 1.0753~\AA\ and an H--C--H angle of 133.82 degrees in
$C_{1}$ symmetry as reported in \cite{kalemos04}. 
The integrals for the CGTN calculations are calculated over the natural
orbitals of the corresponding CASSCF calculation. 
Within this theoretical setup, the CGTN calculations can be seen as 
CASCI-equivalent calculations, where the CI weights are iteratively improved
rather than obtained from an expensive diagonalization step.
For the parallel-tempering Monte Carlo simulation, we use eight replicas at
different temperatures in the range [1$\times 10^{-8}$ $E_h$,0.001 $E_h$].

We investigate three active spaces that are successively enlarged, starting with a 
CAS(4,4) of four spatial orbitals comprising four electrons that is increased in
each step by an occupied and a virtual orbital around the Fermi level yielding 
in total CAS(4,4), CAS(6,6), and CAS(8,8). 
The CAS is specified in parentheses as ($n$,$m$) where
$n$ is the number of electrons in $m$ molecular orbitals.
The number of variational parameters in the CGTN states does not depend on the
dimension of the $N$-particle Hilbert space but on the number of orbitals in
the corresponding active space. The selected active spaces provide insight into
the convergence behavior of the CGTN parametrization. For the CAS(4,4), the number of
variational parameters is around three times larger than the size of the
Hilbert space of CH$_2$ (resulting in an over-parametrization), whereas for CAS(6,6) it is of
comparable size. For the CAS(8,8), however, there are about nine times more many-electron basis
states (i.e., occupation number vectors) than variational parameters in the CGTN ansatz. Hence,
while the first two smaller active spaces allow us to demonstrate that the CGTN ansatz is able
to reproduce the CASSCF reference, the third CAS probes the efficiency of the reduced-parameter CGTN
ansatz.
In order to prevent spin-contamination in the CGTN state, the energy evaluation
is performed in the basis of spin-adapted configuration state functions (CSF)
for the singlet and triplet calculation. The weight for the CSF is
calculated as a linear combination of the weights for the occupation number
vector.

\begin{figure}[H] 
\caption{
Graphical representation of the spin splitting of the singlet and
triplet states of methylene with increasing size of the active space. 
For the singlet ($\times$) and triplet ($\opensquare$) CGTN calculations, a spin-adapted configuration state function
(CSF) basis was employed. The singlet and triplet CASSCF calculations are shown
as ($+$) and ($\opendiamond$), respectively.
\label{fig:spinsplitch2}}
\begin{center}
\vspace*{1.5cm}
\includegraphics[scale=0.5]{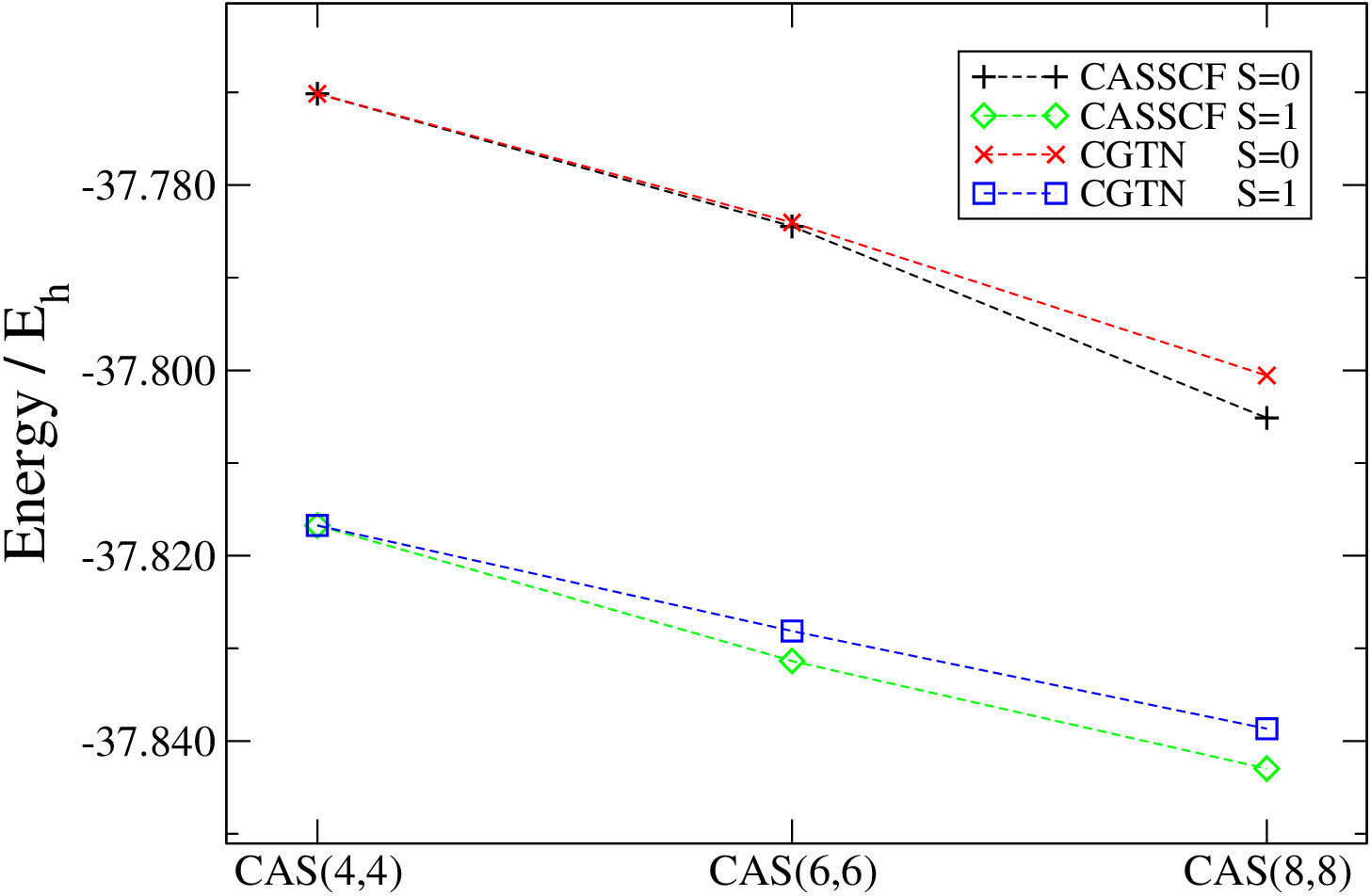}
\end{center}
\end{figure}

In \Fref{fig:spinsplitch2} and \Tref{chtwodata}, 
CASSCF energy splittings between singlet and triplet spin states
are compared to those obtained in the CGTN calculations. The number of variational parameters are given for each
calculation as well. 
The total absolute energies cannot be quantitatively reproduced by the CGTN
states but they provide a qualitatively correct description of the energy
difference between different spin states for a set of active spaces.

\begin{table}[H]
\caption{
Difference of triplet ground state and lowest-lying singlet state energies
of methylene (CH$_2$) as obtained from CASSCF reference and CGTN calculations.
The number of variational parameters in the CASSCF wave function corresponds
to the number of occupation number vectors with the correct particle number and 
projected spin in the given active space: 
$\dim \mathscr{H}_{\rm CASSCF}^{\rm S=0} $
and $\dim \mathscr{H}_{\rm CASSCF}^{\rm S=1} $ parameters.
The singlet and triplet states approximated by the CGTN ansatz comprise an equal number of variational parameters
in the wave functions.
In the last column, the difference between CASSCF and
CGTN spin-splitting energies $\Delta \Delta$E are given in kcal mol$^{-1}$.
\label{chtwodata}}
\begin{center}
\begin{tabular}{lrrrrr}
\br
CAS   & $\Delta$E$_{\rm{CASSCF}}/E_{h}$  & $\dim \mathscr{H}^{\rm S=0} $/
		$\dim \mathscr{H}^{\rm S=1} $ & $\Delta$E$_{\rm{CGTN}} / E_{h}$ &
		$\#{\rm CGTN}$ &
		$\Delta \Delta$E / kcal mol$^{-1}$ \\
\mr
(4,4) & 0.0466 & 36/16     & 0.0466 & 144 &  0.0   \\
(6,6) & 0.0435 & 400/225   & 0.0441 & 312 & $-$0.38 \\
(8,8) & 0.0378 & 4900/3136 & 0.0381 & 544 & $-$0.18 \\
\br
\end{tabular}
\end{center}
\end{table}

For the CAS(4,4), the CGTN calculation exactly reproduce the CASSCF
reference calculations and verify that the ansatz optimized with the 
parallel-tempering Monte Carlo optimization can indeed find the
correct ground-state wave function.
The next question to answer is whether the CGTN ansatz can also extract the
essential features of the
electronic structure for larger active spaces because 
even if total electronic energies are not well reproduced, it would be sufficient to reliably 
produce relative energies of chemical accuracy (of about 1 kcal mol$^{-1}$).
We already found~\cite{marti2008} that MPS as optimized by the DMRG algorithm can reproduce 
the energetical spin splitting in transition metal complexes and clusters although the one-dimensional
MPS parametrization is not very suited for this problem.
The energy difference between two states can converge much faster 
than the total electronic energies of the individual states. 
Considering that during a chemical process (reaction, spin flip) only a small number of orbitals is
needed to qualitatively describe the changes in electronic structure, it can be understood that
the parametrization of the total electronic states requires a balanced representation of all occupation
number vectors that involve these orbitals. We may expect that this balanced description is possible with
a CGTN ansatz. 
This is exactly what we observe in the CAS(6,6)- and CAS(8,8)-CGTN calculations. The relative
energies appear to be better reproduced than the absolute energies for the different spin states.
Even though the parametrization in terms of the CGTN states consists of only a small
fraction of parameters compared to the dimension of the Hilbert space in the CAS(8,8) case, the
vertical spin splitting is accurately reproduced.

\begin{figure}[H] 
\caption{
The squared CI coefficients of the most important occupation number vectors
are shown for the singlet ($+$) and triplet ($\opendiamond$) CASSCF 
and the singlet ($\times$) and triplet ($\opensquare$) CGTN calculations for methylene in a CAS(8,8).
Even though the CGTN state has about 90\% less parameters than the CASSCF 
wave function, it finds the most important occupation number vectors and
provides highly accurate CI coefficients.
\label{fig:ciplot}}
\begin{center}
\vspace*{1.5cm}
\includegraphics[scale=0.5]{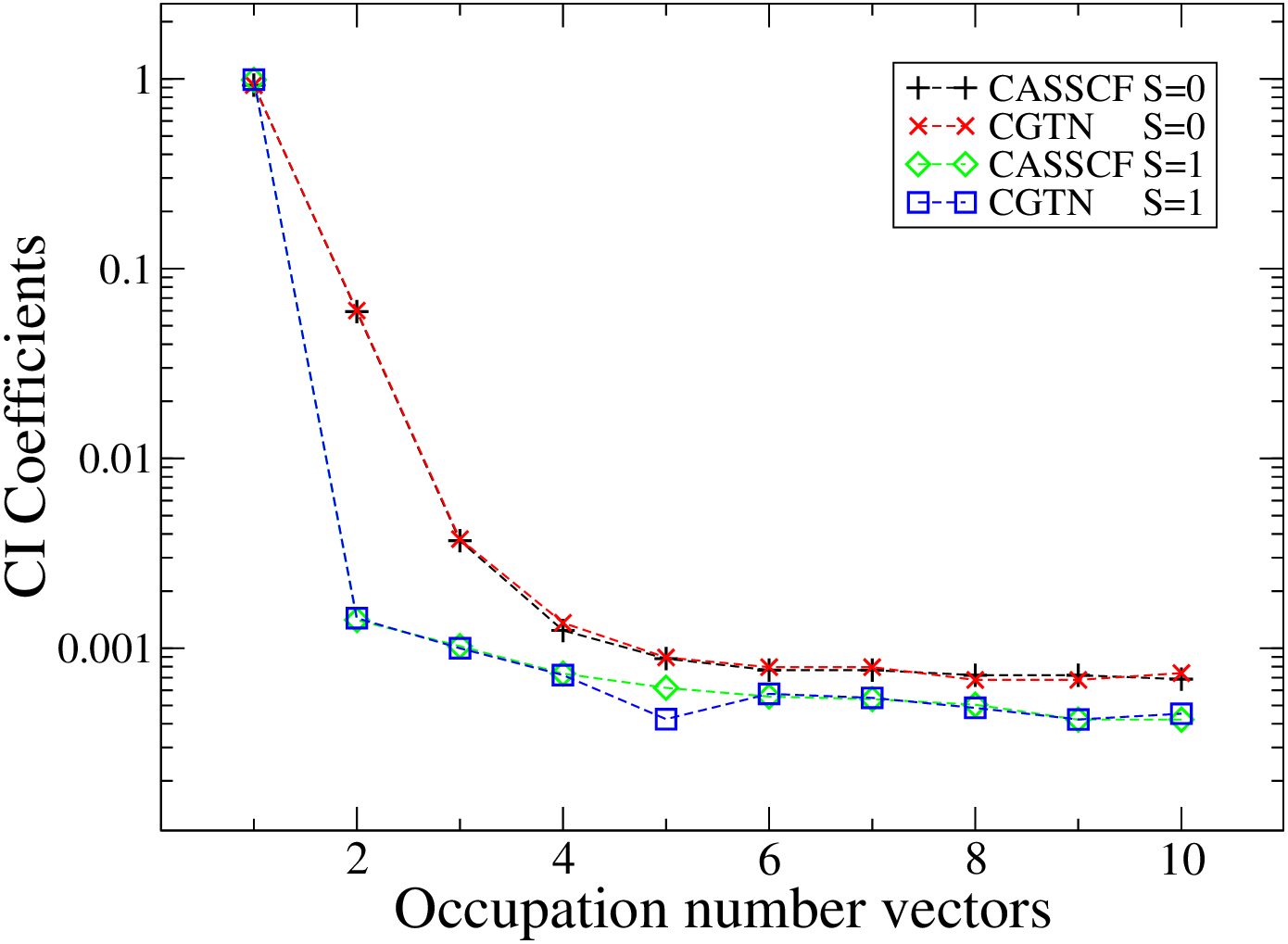}
\end{center}
\end{figure}

The accuracy of the CGTN to represent the electronic structure can also be
assessed by performing an analysis of the CI coefficients of the CASSCF and the
CGTN calculations. In \Fref{fig:ciplot} and \Tref{citable}, the CI
coefficients of the ten most important occupation number vectors
for the singlet and triplet CGTN and CASSCF calculations are compared. 

\begin{table}[H]
\caption{
The CI coefficients of the CASSCF 
and the CGTN wave functions of singlet and
triplet spin symmetry are shown for the
ten most important occupation number vectors. The CGTN coefficients 
qualitatively and even quantitatively 
agree with the CASSCF 
reference which is the exact solution for the CAS(8,8) in the
given one-particle basis. 
\label{citable}}
\begin{center}
\begin{tabular}{lrrrr}
\br
ONV  &  C$_{\rm CASSCF}^{\rm (S=0)}$ &  C$_{\rm CGTN}^{\rm (S=0)}$ & 
                 C$_{\rm CASSCF}^{\rm (S=1)}$ &  C$_{\rm CGTN}^{\rm (S=1)}$ \\
\mr
 1  &   0.9623   &   0.9624   &    0.9945  &    0.9950  \\        
 2  &  -0.2436   &  -0.2452   &   -0.0375  &   -0.0380  \\
 3  &  -0.0607   &  -0.0613   &   -0.0320  &   -0.0316  \\
 4  &  -0.0352   &  -0.0369   &   -0.0271  &   -0.0269  \\
 5  &  -0.0297   &  -0.0299   &    0.0249  &    0.0206  \\
 6  &  -0.0277   &  -0.0282   &    0.0236  &    0.0240  \\
 7  &  -0.0277   &  -0.0282   &    0.0232  &    0.0234  \\
 8  &   0.0269   &   0.0261   &   -0.0225  &   -0.0220  \\
 9  &   0.0269   &   0.0261   &   -0.0205  &   -0.0205  \\
10  &  -0.0262   &  -0.0272   &    0.0205  &    0.0213  \\
\br
\end{tabular}
\end{center}
\end{table}

Qualitatively speaking, the CGTN states ``carved" the important occupation number vectors
out of the entire $N$-electron Hilbert space 
that characterizes the electronic structure of the underlying
molecular system. Therefore, a qualitatively correct description of
the electronic structure is provided by the CGTN wave function.

\subsection{Strongly Correlated Molecular System: Ozone\label{ozone}}

We continue our numerical study with a most difficult case selected to probe the capabilities of
the CGTN ansatz: The electronic structure of ozone at the transition state structure of the
O$_2$ + O chemical reaction is a complex multireference problem~\cite{schinke2004}.
We performed CAS(8,9)-CI reference calculations for the singlet and triplet
states of ozone at the transition structure reported in \cite{schinke2004}.
For this calculation, we select Dunning's cc-pVTZ basis set~\cite{dunn1989,dunn1992} and an
active space consisting of the $9-14a^\prime\ 1-3a^{\prime \prime}$ molecular orbitals.
The Hilbert space of the singlet state is then spanned by 7'956 
and the one for the triplet state by 5'268 occupation number vectors.
The CGTN state contains only 684 variational parameters 
which is an efficient reduction by 91\% compared to the singlet CASSCF wave function.
The singlet energies of the ozone molecule at a transition state structure are
given in \Tref{ozoneenergies} and show
the highly multi-reference nature of the electronic structure 
when compared to the Hartree--Fock energy.

\begin{table}[H]
\caption{
Electronic energy of the singlet state of transition-state ozone in Hartree.
The correlation energy ($E_{corr}$) denotes the energy difference between the
Hartree--Fock energy and the energy obtained by a correlation method.
All calculations were performed with an active space of 9 molecular orbitals comprising 8
electrons.
The DMRG calculations were taken from \cite{mori07} for comparison and
$m$ represents the number of many-particle DMRG system basis states.
The CGTN energy for the singlet state is evaluated in a spin-adapted configuration
state function (CSF) basis.
	\label{ozoneenergies}}
\begin{center}
\begin{tabular}{lrr}
\br
	Method & E$_{\rm S=0}$ / $E_h$ & $E_{corr}$\\
\mr
	HF	   		& $-$224.282 841 & 0\%  \\
	CASCI       & $-$224.384 301 & 100\% \\
	DMRG(m=48)  & $-$224.384 252 & 99\% \\
	DMRG(m=156) & $-$224.384 301 & 100\% \\
	CGTN        & $-$224.381 648 & 97\% \\
\br
\end{tabular}
\end{center}
\end{table}

The vertical spin splitting between the first excited
singlet and ground-state triplet state is reported in \Tref{spinozone}.
For the first excited state of singlet symmetry a CSF basis has been constructed
to obtain a spin pure state withouth spin contamination. 
For the triplet ground state, we have
tested the levelshift approach as well as no spin constrains at all and
found that levelshift calculations are prone to get
stuck in a local minimum. This can be circumvented if no spin constraints 
are applied. Then, however, spin contamination might
become a problem. For the levelshift calculation, $\epsilon=1$ was used.
As in the previous study of methylene, the relative energies between the
singlet and triplet states converge much faster
than the total electronic energies  --- even for this highly multireference
system. (Note that we used the same temperature range and number of replicas as
in the case of methylene.)

\begin{table}[H]
\caption{
Vertical spin-splitting energy differences between the singlet and triplet state of ozone
at the transition geometry.
All calculations were in an active space of 9 molecular orbitals comprising 8
electrons.
For the singlet CGTN energy, a CSF basis was employed whereas 
no spin constraints were imposed on the triplet state calculation.
\label{spinozone}}
\begin{center}
\begin{tabular}{lrrr}
\hline \hline
& E$_{\rm{S=0}}$/$E_h$ & E$_{\rm{S=1}}$/$E_h$ &
$\Delta$E/kcalmol$^{-1}$ \\
\hline
HF    & -224.282 841 & -224.357 167 & 46.640 \\
CASCI & -224.384 301 & -224.416 172 & 19.999 \\
CGTN  & -224.381 648 & -224.412 775 & 19.532 \\
\hline \hline
\end{tabular}
\end{center}
\end{table}

\begin{figure}[H] 
\caption{
The convergence of the energy for the replica with $T=1.0\times10^{-8}$ E$_h$ of
	the triplet CGTN state of ozone in an
active space of 9 molecular orbitals and 8 electrons is shown. The inner panel shows a
zoom of the first thousands Monte Carlo iterations with the Hartree--Fock and
CASCI energies.
It is evident that the calculation employing the levelshift operator
got stuck
in a local minimum. The other CGTN calculation has no spin restriction. 
However, since the ground state is a triplet spin state, no spin contamination
is expected. This is also seen in the expectation value of the 
converged triplet CGTN state of $\langle{\hat{S}^2}\rangle = 1.99$.
\label{fig:o3tripconv}}
\begin{center}
\vspace*{1.5cm}
\includegraphics[scale=0.5]{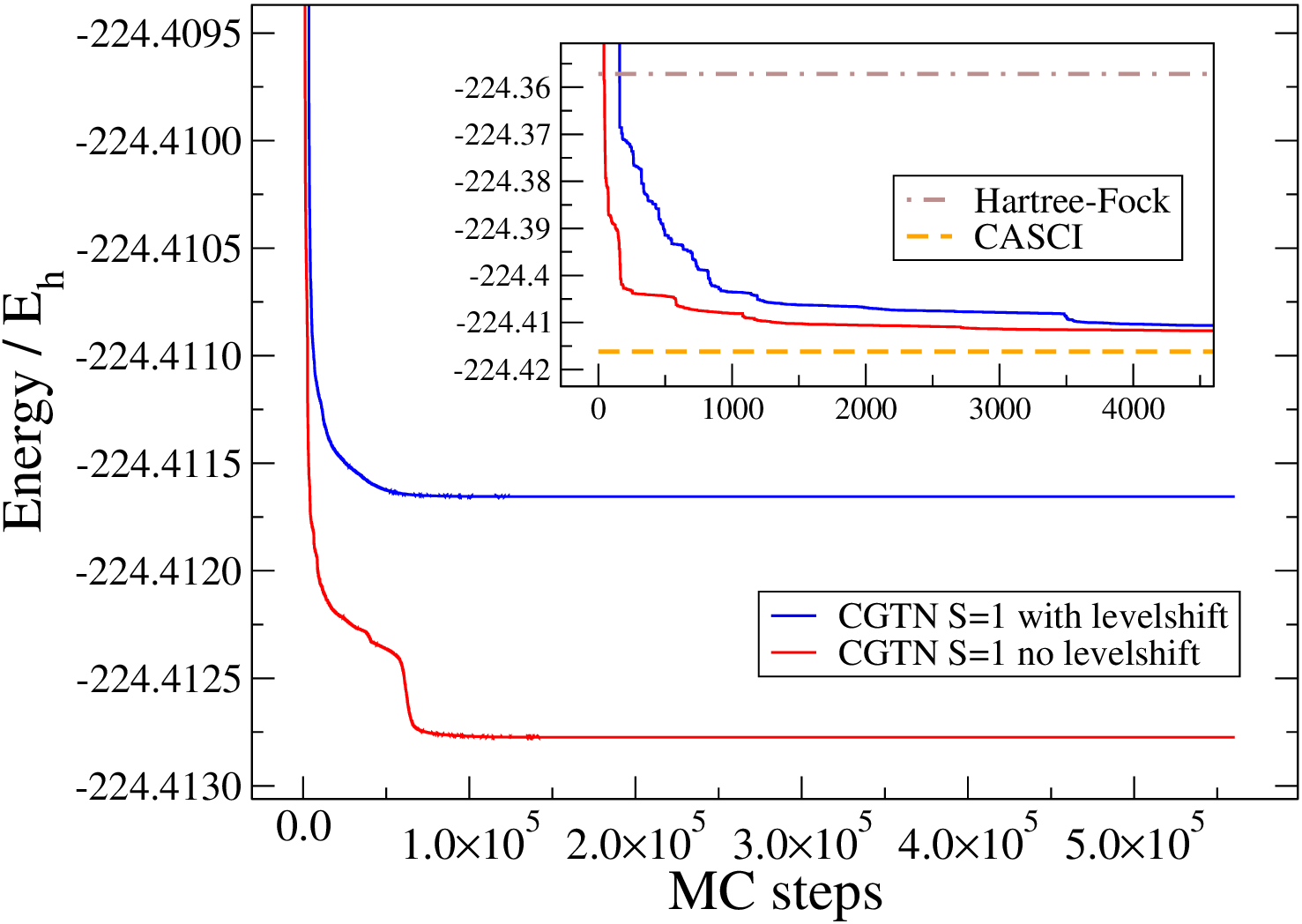}
\end{center}
\end{figure}

In \Fref{fig:o3tripconv}, the convergence behaviour for the triplet ground
state calculation is shown. It can be seen that convergence difficulties arise when 
the levelshift operator is applied.
In the algorithm, the $\hat{S}_{-}\hat{S}_{+}$ operator translates into a
high-order polynomial which features 
many roots and therefore many local minima.

\section{Conclusions\label{conclusions}}
In this work, we introduced a general class of tensor network states, which we denoted Complete-Graph Tensor Network states,
to approximate the electronic wave function of a molecular system. This ansatz assumes pair correlations of one-electron
states (orbitals) to construct {\it all} CI expansion coefficients of a total electronic state. Hence, instead of
$2^k$ --- or $4^k$ in the case of spatial orbitals --- variational parameters,
we employ only $k(k+1)/2\times q^2$.
CGTN states are a subclass of the most general CPS form of tensor network approximations to the FCI state.
The accuracy of the CGTN approximation of total electronic states of molecules has been numerically studied for methylene and ozone.
We should note that this is the first application of a tensor network parametrization for molecular wave functions 
employing the full non-local electronic many-particle Hamiltonian.
For this purpose, the $k(k+1)/2\times q^2$ CGTN parameters have been optimized with a variational Monte Carlo protocol that we have augmented by 
parallel tempering in order to prevent convergence to local minima of the electronic energy hypersurface in this
parameter space.

In molecular physics and chemistry, we are primarily interested in obtaining accurate relative energies
between two spin states or between two configurations on the same potential energy surface 
to describe the thermodynamics and kinetics of chemical reactions. The accurate calculation of 
relative energies is therefore mandatory. CGTN states provide the flexibility to
describe electronic structures without relying on an {\it a priori} chosen
reference configuration such as the Hartree--Fock state. The CGTN ansatz is
therefore not biased to any particular Slater determinant and capable of finding
the most important occupation number vectors in the Hilbert space of the molecular system.

\ack{Acknowledgments}
KM and MR gratefully acknowledge financial support 
through a TH-Grant (TH-26 07-3) from ETH Zurich and
FV by the EU (ERC grant QUERG).
Parts of this work have been carried out at the Institute of Pure and Applied
Mathematics, UCLA and the Erwin Schr\"odinger Institute, Vienna.

\section*{References}

\bibliographystyle{unsrt}

\begin{thebibliography}{10}

\bibitem{whit92a}
White S R.
\newblock {Density matrix formulation for quantum renormalization groups}.
\newblock {\em Phys. Rev. Lett.}, 69(19):2863--2866, 1992.

\bibitem{whit92b}
White S R and Noack R M.
\newblock {Real-space quantum renormalization groups}.
\newblock {\em Phys. Rev. Lett.}, 68(24):3487--3490, 1992.

\bibitem{whit93}
White S R.
\newblock {Density-matrix algorithms for quantum renormalization groups}.
\newblock {\em Phys. Rev. B}, 48(14):10345--10356, 1993.

\bibitem{scholl05}
Schollw\"ock U.
\newblock {The density-matrix renormalization group}.
\newblock {\em Rev. Mod. Phys.}, 77:259--315, 2005.

\bibitem{chan08}
Chan G K-L, Dorando J J, Ghosh D, Hachmann J,
  Neuscamman E, Wang H and Yanai T.
\newblock {An introduction to the density matrix renormalization group ansatz
  in quantum chemistry}.
\newblock In {Wilson S, Grout P J, Maruani J, DelgadoBarrio G and
  Piecuch P},\ (Ed.), {\em {Frontiers in Quantum Systems in Chemistry and
  Physics}}, volume~{18} of {\em {Prog. Theor. Chem. Phys.}}, pages {49--65},
  {2008}.

\bibitem{mart09}
Marti K H and Reiher M.
\newblock {The density matrix renormalization group algorithm in quantum
  chemistry}.
\newblock {\em Z. Phys. Chem.}, pages DOI:10.1524, in press., 2009.

\bibitem{rommerostlund1995}
{\"O}stlund S and Rommer S.
\newblock {Thermodynamic limit of density-matrix renormalization}.
\newblock {\em {Phys. Rev. Lett.}}, {75}({19}):{3537--3540}, {1995}.

\bibitem{rommerostlund1997}
Rommer S and {\"O}stlund S.
\newblock {Class of ansatz wave functions for one-dimensional spin systems and
  their relation to the density matrix renormalization group}.
\newblock {\em {Phys. Rev. B}}, {55}({4}):{2164--2181}, {1997}.

\bibitem{fannes1992}
Fannes M, Nachtergaele B and Werner R F.
\newblock {Finitely correlated states on quantum spin chains}.
\newblock {\em {Commun. Math. Phys.}}, {144}({3}):{443--490}, {1992}.

\bibitem{fannes1992b}
Fannes M, Nachtergaele B and Werner R F.
\newblock {Abundance of translation invariant pure states on quantum spin
  chains}.
\newblock {\em {Lett. Math. Phys.}}, {25}({3}):{249--258}, {JUL} {1992}.

\bibitem{aklt}
Affleck I, Kennedy T, Lieb E H and Tasaki H.
\newblock {Valence bond ground-states in isotropic quantum antiferromagnets}.
\newblock {\em {Commun. Math. Phys.}}, {115}({3}):{477--528}, {1988}.

\bibitem{verstraete2008}
Verstraete F, Murg V and Cirac J I.
\newblock {Matrix product states, projected entangled pair states, and
  variational renormalization group methods for quantum spin systems}.
\newblock {\em {Adv. Phys.}}, {57}({2}):{143--224}, {2008}.

\bibitem{reih05a}
Moritz G, Hess B A and Reiher M.
\newblock {Convergence behavior of the density-matrix renormalization group
  algorithm for optimized orbital orderings}.
\newblock {\em J. Chem. Phys.}, 122(2):024107, 2005.

\bibitem{reih06}
Moritz G and Reiher M.
\newblock {Construction of environment states in quantum-chemical
  density-matrix renormalization group calculations}.
\newblock {\em J. Chem. Phys.}, 124(3):034103, 2006.

\bibitem{meinbuch}
Reiher M and Wolf A.
\newblock {\em {Relativistic Quantum Chemistry}}.
\newblock Wiley-VCH, Weinheim, 2009.

\bibitem{verstraete2009}
Cirac J I and Verstraete F.
\newblock {Renormalization and tensor product states in spin chains and
  lattices}, 2009.
\newblock arXiv.org:0910.1130.

\bibitem{vidal2003a}
Vidal G.
\newblock {Efficient Classical Simulation of Slightly Entangled Quantum
  Computations}.
\newblock {\em Phys. Rev. Lett.}, 91(14):147902, 2003.

\bibitem{eisert2008}
Eisert J, Cramer M and Plenio M B.
\newblock Area laws for the entanglement entropy - a review.
\newblock {\em Preprint}, 2008.
\newblock arXiv:0808.3773v3.

\bibitem{hastings2006}
Hastings M B.
\newblock {Solving gapped Hamiltonians locally}.
\newblock {\em Phys. Rev. B}, 73:085115, 2006.

\bibitem{helgaker_book}
Helgaker T, J{\o}rgensen P and Olsen J.
\newblock {\em {Molecular Electronic-Structure Theory}}.
\newblock John Wiley \& Sons Ltd., Chichester, 2000.

\bibitem{sierra1998}
Sierra G and Martin-Delgado M A.
\newblock {The Density Matrix Renormalization Group, Quantum Groups and
  Conformal Field Theory}.
\newblock {\em Preprint}, 1998.
\newblock arXiv:cond-mat/9811170v3.

\bibitem{nishino2001}
Nishino T, Hieida Y, Okunushi K, Maeshima N, Akutsu Y and Gendiar A.
\newblock {Two-Dimensional Tensor Product Variational Formulation}.
\newblock {\em Progr. Theor. Phys.}, 105(3):409--417, 2001.

\bibitem{nishio2004}
Nishio Y, Maeshima N, Gendiar A and Nishino T.
\newblock {Tensor Product Variational Formulation for Quantum Systems}.
\newblock {\em Preprint}, 2004.
\newblock arXiv:cond-mat/0401115v1.

\bibitem{verstraete2004}
Verstraete F, Garcia-Ripoll J J and Cirac J I.
\newblock {Matrix product density operators: Simulation of finite-temperature
  and dissipative systems}.
\newblock {\em {Phys. Rev. Lett.}}, {93}({20}):{207204}, {2004}.

\bibitem{martindelgado2002}
Martin-Delgado M A, Rodriguez-Laguna J and Sierra G.
\newblock {Density-matrix renormalization-group study of excitons in
  dendrimers}.
\newblock {\em {Phys. Rev. B}}, {65}({15}):{155116}, {2002}.

\bibitem{verstraete2006}
Verstraete F, Wolf M M, Perez-Garcia D and Cirac J I.
\newblock {Criticality, the area law, and the computational power of projected
  entangled pair states}.
\newblock {\em {Phys. Rev. Lett.}}, {96}({22}):{220601}, {2006}.

\bibitem{zhou2008}
Zhou H Q, Orus R and Vidal G.
\newblock {Ground state fidelity from tensor network representations}.
\newblock {\em {Phys. Rev. Lett.}}, {100}({8}):{080601}, {2008}.

\bibitem{shi2006}
Shi Y Y, Duan L M and Vidal G.
\newblock {Classical simulation of quantum many-body systems with a tree tensor
  network}.
\newblock {\em {Phys. Rev. A}}, {74}({2}):{022320}, {2006}.

\bibitem{vidal2007}
Vidal G.
\newblock {Entanglement renormalization}.
\newblock {\em {Phys. Rev. Lett.}}, {99}({22}):{220405}, {2007}.

\bibitem{perez2007}
Perez-Garcia D, Verstraete F, Wolf M M and Cirac J I.
\newblock {Matrix product state representations}.
\newblock {\em {Quantum Inf. Comput.}}, {7}({5-6}):{401--430}, {2007}.

\bibitem{klumper1991}
Kl{\"u}mper A, Schadschneider A and Zittartz J.
\newblock {Equivalence and solution of anisotropic spin-1 models and
  generalized T-J fermion models in one dimension}.
\newblock {\em {J. Phys. A: Math. Gen.}}, {24}({16}):{L955--L959}, {1991}.

\bibitem{klumper1992}
Kl{\"u}mper A, Schadschneider A and Zittartz J.
\newblock {Ground-state properties of a generalized VBS-model}.
\newblock {\em {Z. Phys. B}}, {87}({3}):{281--287}, {1992}.

\bibitem{derrida1993}
Derrida B, Evans M R, Hakim V and Pasquier V.
\newblock {Exact solution of a 1D asymmetric exclusion model using a matrix
  formulation}.
\newblock {\em {J. Phys. A: Math. Gen.}}, {26}({7}):{1493--1517}, {1993}.

\bibitem{mccul2007}
McCulloch I P.
\newblock {From density-matrix renormalization group to matrix product states}.
\newblock {\em {J. Stat. Mech: Theory Exp.}}, pages {1742--5468}, {2007}.

\bibitem{chan06}
Hachmann J, Cardoen W and Chan G K-L.
\newblock {Multireference correlation in long molecules with the quadratic
  scaling density matrix renormalization group}.
\newblock {\em J. Chem. Phys.}, 125(14):141101, 2006.

\bibitem{schuch2008}
Schuch N, Wolf M M, Verstraete F and Cirac J I.
\newblock {Simulation of quantum many-body systems with strings of operators
  and Monte Carlo tensor contractions}.
\newblock {\em {Phys. Rev. Lett.}}, {100}({4}):{040501}, {FEB 1} {2008}.

\bibitem{mezzacapo2009}
Mezzacapo F, Schuch N, Boninsegni M and Cirac J I.
\newblock {Ground-state properties of quantum many-body systems:
  entangled-plaquette states and variational Monte Carlo}.
\newblock {\em New Journal of Physics}, 11:083026, 2009.
\newblock arXiv.org:0905.3898.

\bibitem{chan09}
Changlani H J, Kinder J M, Umrigar C J and Chan G K-L.
\newblock Approximating strongly correlated spin and fermion wavefunctions with
  correlator product states, 2009.
\newblock arXiv.org:0907.4646.

\bibitem{huse1988}
Huse D A and Elser V.
\newblock {Simple variational wave-functions for two-dimensional heisenberg
  spin-1/2 antiferromagnets}.
\newblock {\em {Phys. Rev. Lett.}}, {60}({24}):{2531--2534}, {1988}.

\bibitem{sandvik2007}
Sandvik A W and Vidal G.
\newblock {Variational quantum Monte Carlo simulations with tensor-network
  states}.
\newblock {\em {Phys. Rev. Lett.}}, {99}({22}):{220602}, {NOV 30} {2007}.

\bibitem{fren02}
Frenkel D and Smit B.
\newblock {\em {Understanding Molecular Simulation}}.
\newblock Academic Press, San Diego, 2nd edition, 2002.

\bibitem{katzgraber2006}
Katzgraber H G, Trebst S, Huse D A and Troyer M.
\newblock Feedback-optimized parallel tempering monte carlo.
\newblock {\em {J. Stat. Mech: Theory Exp.}}, 2006(03):P03018, 2006.

\bibitem{paul01}
Paulsen H, Duelund L, Winkler H, Toftlund H and Trautwein A X.
\newblock {Free energy of spin-crossover complexes calculated with density
  functional methods}.
\newblock {\em Inorg. Chem.}, 40:2201--2203, 2001.

\bibitem{reih01}
Reiher M, Salomon O and Hess B A.
\newblock {Reparameterization of hybrid functionals based on energy differences
  of states of different multiplicity}.
\newblock {\em Theor. Chem. Acc.}, 107(1):48, 2001.

\bibitem{reih02h}
Reiher M.
\newblock {Theoretical study of the [Fe(phen)$_2$(NCS)$_2$] spin-crossover
  complex with reparametrized density functionals}.
\newblock {\em Inorg. Chem.}, 41:6928--6935, 2002.

\bibitem{salo02}
Salomon O, Reiher M and Hess B A.
\newblock {\em J. Chem. Phys.}, 117:4729--4737, 2002.

\bibitem{poli03}
Poli R and Harvey J N.
\newblock {Spin forbidden chemical reactions of transition metal compounds. New
  ideas and new computational challenges}.
\newblock {\em Chem. Soc. Rev.}, 32:1--8, 2003.

\bibitem{harv04}
Harvey J N.
\newblock {DFT computation of relative spin-state energetics of transition
  metal compounds}.
\newblock {\em Struct. Bonding}, 112:151--183, 2004.

\bibitem{reih06_fd135}
Reiher M.
\newblock {On the definition of local spin in relativistic and nonrelativistic
  quantum chemistry}.
\newblock {\em Faraday Discuss.}, 135:97--124, 2007.

\bibitem{schr00}
Schr{\"o}der D, Shaik S and Schwarz H.
\newblock {Two-State Reactivity as a New Concept in Organometallic Chemistry}.
\newblock {\em Acc. Chem. Res.}, 33(3):139--145, 2000.

\bibitem{shaik1999}
Harris N, Shaik S, Schr{\"o}der D and Schwarz H.
\newblock {Single- and two-state reactivity in the gas-phase C-H band
  activation of norbarnane by `bare' FeO+}.
\newblock {\em {Helv. Chim. Acta}}, {82}({10}):{1784--1797}, {1999}.

\bibitem{marti2008}
Marti K H, Malkin Ondik I, Moritz G and Reiher M.
\newblock {Density matrix renormalization group calculations on relative
  energies of transition metal complexes and clusters}.
\newblock {\em {J. Chem. Phys.}}, {128}({1}):{014104}, {JAN 7} {2008}.

\bibitem{kalemos04}
Kalemos A, Dunning~Jr T H, Mavridis A and Harrison J F.
\newblock {CH$_{\textrm{2}}$ revisited}.
\newblock {\em Canadian Journal of Chemistry}, 82(6):684--693, 2004.

\bibitem{molpro2002.6}
{Werner H J, Knowles P J, Lindh R, Sch\"{u}tz M {\em et al.}}
\newblock {\sc Molpro}, version 2002.6, a package of ab initio programs, 2002.
\newblock See http://www.molpro.net.

\bibitem{dunn1989}
{Dunning Jr.} T H.
\newblock {Gaussian basis sets for use in correlated molecular calculations. I.
  The atoms boron through neon and hydrogen}.
\newblock {\em J. Chem. Phys.}, 90:1007, 1989.

\bibitem{dunn1992}
Kendall R A, {Dunning Jr.} T A and Harrison R J.
\newblock {Electron affinities of the first row atoms revisited. Systematic
  basis sets and wave functions}.
\newblock {\em J. Chem. Phys.}, 96:6796, 1992.

\bibitem{schinke2004}
Schinke R and Fleurat-Lessard P.
\newblock {The transition-state region of the O(P-3)+O-2((3)Sigma(-)(g))
  potential energy surface}.
\newblock {\em {J. Chem. Phys.}}, {121}({12}):{5789--5793}, {2004}.

\bibitem{mori07}
Moritz G, Marti K H and Reiher M.
\newblock ``{{\sc Qc-Dmrg-ETH} 2.0, A program for quantum chemical DMRG
  calculations}''.
\newblock Copyright 2007--2009, ETH Zurich.

\end{thebibliography}

\end{document}